\documentclass[12pt]{article}

\usepackage{epsfig}				
\usepackage{amssymb,amsmath}
\usepackage{multirow}
\usepackage{rotating}
\usepackage{graphicx}
\usepackage{lscape}
\usepackage{hyperref}

\usepackage{calrsfs}		

\newcommand{\bea}{\begin{eqnarray}}
\newcommand{\eea}{\end{eqnarray}}

 \makeatletter
\def\alt{\mathrel{\mathpalette\gl@align<}}
\def\agt{\mathrel{\mathpalette\gl@align>}}
\def\gl@align#1#2{\lower.6ex\vbox{\baselineskip\z@skip\lineskip\z@
\ialign{$\m@th#1\hfil##\hfil$\crcr#2\crcr\sim\crcr}}} \makeatother

\begin{document}

\begin{flushright}
\end{flushright}

\vspace*{1.0cm}

\begin{center}
\baselineskip 20pt 
{\Large\bf 
125 GeV Higgs boson mass and muon $g-2$ in 5D MSSM
}
\vspace{1cm}

{\large 
Nobuchika Okada$^{a,}$\footnote{ E-mail: okadan@ua.edu} 
and 
Hieu Minh Tran$^{a,b,}$\footnote{ E-mail: hieu.tranminh@hust.edu.vn} 
} 
\vspace{.5cm}

{\baselineskip 20pt \it
$^a$Department of Physics and Astronomy, University of Alabama, \\ 
Tuscaloosa, Alabama 35487, USA \\
\vspace{2mm} 
$^b$Hanoi University of Science and Technology, 
1 Dai Co Viet Road, Hanoi, Vietnam \\
}

\vspace{.5cm}

\vspace{1.5cm} {\bf Abstract}
\end{center}

In the MSSM, the tension between the observed Higgs boson mass and 
   the experimental result of the muon $g-2$ measurement requires a large mass 
   splitting between stops and smuons/charginos/neutralinos.   
We consider a 5-dimensional (5D) framework of the MSSM with the Randall-Sundrum warped background metric, 
   and show that such a mass hierarchy is naturally achieved in terms of geometry. 
In our setup, the supersymmetry is broken at the ultraviolet (UV) brane, while all the MSSM multiplets reside 
   in the 5D bulk. 
An appropriate choice of the bulk mass parameters for the MSSM matter multiplets 
  can naturally realize the sparticle mass hierarchy desired to resolve the tension. 
Gravitino is localized at the UV brane and hence becomes very heavy, 
  while the gauginos spreading over the bulk acquire their masses 
  suppressed by the 5th dimensional volume.   
As a result, the LSP neutralino is a candidate for the dark matter as usual in the MSSM. 
In addition to reproducing the SM-like Higgs boson mass of around 125 GeV and 
  the measured value of the muon $g-2$, we consider a variety of phenomenological constraints, 
  and present the benchmark particle mass spectra 
  which can be explored at the LHC Run-2 in the near future.

\thispagestyle{empty}

\newpage

\addtocounter{page}{-1}

\baselineskip 18pt

\section{Introduction}  


The standard model (SM) of elementary particles has been tested to a very high accuracy.
It can be considered to be completed in the sense that all particles in the SM have been observed 
  and their properties have been confirmed to be consistent with the SM expectations. 
However, there are still problems that this model itself cannot address, and hence motivating us to go beyond the SM.
The discovery of the Higgs boson not only fulfills the SM particle content, but also provides us with a hint 
   that the SM needs to be extended according to the requirement of naturalness.
Cosmological observations confirmed the existence of dark matter that goes beyond the SM prediction.
On the other hand, the measurement for the muon anomalous magnetic dipole moment ($g-2$) \cite{Exp_g-2}
   reveals a 3-4 $\sigma$ discrepancy between the measured central value and the SM prediction \cite{SM_g-2}.


Supersymmetry (SUSY) has been investigated for a long time as one of the most promising candidates beyond the SM. 
The gauge hierarchy problem, in other words, the instability of the electroweak scale under quantum corrections can be 
   solved by a SUSY extension of the SM such as the minimal supersymmetric SM (MSSM). 
Assuming the conservation of R-parity, SUSY models can provide good candidates for the cold dark matter in the universe.
Additionally, the muon $g-2$ also receives contributions from superpartners (smuons, charginos and neutralinos) and 
  shifts to the allowed measured interval when their masses lie at the electroweak scale \cite{SUSY(g-2)enhance}.


The Higgs boson mass measurement at the Large Hadron Collider (LHC) together with other experiments have put severe constraints on SUSY breaking parameters.
At the tree level, the SM-like Higgs boson mass $m_h$ is just about the Z boson mass $m_Z$. 
To reproduce the Higgs boson mass of about 125 GeV \cite{mHiggs}, quantum corrections to the Higgs boson mass play a crucial role.
The approximate formula for $m_h$ with radiative corrections 
  (in case $A_t$ is relatively large compared to $\mu$, and $m_A \gg m_Z$) is given by \cite{mHapproximation}
\begin{eqnarray}
m_h^2	& \simeq &	m_Z^2 \cos^2 2 \beta 
			+ \frac{3}{4\pi^2} y_t^2 m_t^2 \sin^2 \beta  
			\left[ \log 
					\left( \frac{m^2_{\tilde{t}}}{m_t^2} 
					\right) 
					+ \frac{X_t^2}{m^2_{\tilde{t}}}
					- \frac{X_t^4}{12 m^4_{\tilde{t}}}
			\right],
\label{Higgs}
\end{eqnarray}
where $X_t = A_t - \mu \cot \beta$ is the stop mixing parameter.
In this formula, a large stop mass plays a crucial role to push up the Higgs boson mass 
   from $m_Z$ to the measured value of about 125 GeV.
In many SUSY breaking models, heavy stops imply that other sfermions are also heavy, 
   such that squark masses are of $\mathcal{O}(10\text{TeV})$, and slepton masses lie around a few TeV \cite{Okada:2012nr}.
However, heavy smuons make their loop contributions to the muon anomalous magnetic moment, $a_\mu=\frac{1}{2}(g_\mu-2)$, 
   too small to explain the 3-4 $\sigma$ discrepancy.
This fact can be easily seen in the formula of SUSY contributions to the muon anomalous magnetic moment \cite{amu-formula}:
\begin{eqnarray}
\Delta a_\mu &=&
	\frac{\alpha m_\mu^2 \, \mu M_2 \tan \beta}{4 \pi \sin^2 \theta_W m^2_{{\tilde{\mu}}_L}}
	\left[
		\frac{f_\chi (M_2^2 / m^2_{{\tilde{\mu}}_L}) - f_\chi (\mu^2 / m^2_{{\tilde{\mu}}_L})}			{M_2^2 - \mu^2}
	\right]  			\nonumber \\ 
 & &
 	+ \; \frac{\alpha m_\mu^2 \, \mu M_1 \tan \beta}{4 \pi \cos^2 \theta_W
 		(m^2_{{\tilde{\mu}}_R} - m^2_{{\tilde{\mu}}_L})}
	\left[
		\frac{f_N (M_1^2 / m^2_{{\tilde{\mu}}_R})}{m^2_{{\tilde{\mu}}_R}} -
		\frac{f_N (M_1^2 / m^2_{{\tilde{\mu}}_L})}{m^2_{{\tilde{\mu}}_L}}
	\right],
\label{Delta-a_mu}
\end{eqnarray}
where the loop functions are defined as 
\begin{eqnarray}
f_\chi (x) = \frac{x^2 - 4x + 3 + 2 \ln x}{(1-x)^3} , \;   \; \; 
f_N (x) &=& \frac{x^2 - 1 - 2x \ln x}{(1-x)^3}. 
\end{eqnarray}
For $x={\cal O}(1)$, they are of order one, for example, $f_\chi (1) = -\frac{2}{3}$ and $f_N (1) = -\frac{1}{3}$. 


To yield $\Delta a_\mu \sim 10^{-9}$ to fill the discrepancy between the experimental result 
    and the SM prediction, light smuons and charginos/neutralinos are necessary, 
    while a large $\tan \beta$ works to enhance $\Delta a_\mu$.
A solution to the tension between the Higgs boson mass and the muon $g-2$ may come from a large mass splitting 
   between stops (squarks in general) and smuons (leptons in general)/charginos/neutralinos \cite{mass_splitting}. 
According to Eq.~(\ref{Higgs}), beside heavy stops the Higgs boson mass can also be improved 
   by large $X_t$ \cite{Example_Xt-enhancement, Babu:2014lwa,LargeAt}.
There are also other proposals to solve this tension \cite{Other_g-2}.


In this paper, we investigate the MSSM in the 5D space-time 
    with the Randall-Sundrum (RS) background metric \cite{Randall:1999ee}.
Originally, the RS model was proposed to solve the gauge hierarchy problem of the SM, 
    where all the SM particles are confined on the so-called infrared (IR) brane 
    at a fixed point of the $S^1/Z_2$ orbifold on which the 5th dimension is compactified. 
The large hierarchy between the electroweak and the Planck scales is naturally generated 
    via the so-called warp factor induced by the RS warped background metric.
Soon after the original work, the RS model was extended to have the SM fields to reside in the bulk 
    while the SM Higgs field is confined on the IR brane to maintain the solution to the gauge hierarchy problem \cite{bulkSM}.
SUSY extensions of the RS model then came up with the component field formulation \cite{componentSUSY-RS} 
  and the superfield formulation \cite{superfield-RS}.
This context provides not only an elegant explanation of the diversity of particle masses \cite{Huber:2000ie}, 
    but also a variety of possibilities for SUSY breaking mediation mechanisms with the 5th dimension.%
\footnote{Beside the RS scenario, the flat extra dimension scenario also brings an interesting landscape for discussions on SUSY breaking. See, for example, \cite{largeY} and references therein. }
In SUSY RS models, the SUSY can solve the gauge hierarchy problem as usual, so that a very strong warp factor 
    is not necessary. 
SUSY RS models have the ability to simultaneously solve both hierarchy problems: 
    the gauge hierarchy and the fermion mass (Yukawa) hierarchy.
On the other hand, the AdS/CFT correspondence brings an interesting point to our study setup.
This conjecture maps the physics in the AdS$_5$ space on to its dual 4D picture 
    with a strongly coupled conformal field theory \cite{Gherghetta:2000kr}.
A connection of the models to string theories would be possible \cite{string}.

Our setup is similar to Ref. \cite{Okada:2011ed} that all matter and gauge superfields reside in the bulk. 
But we arrange the Higgs superfields to reside in the bulk as well, while the SUSY breaking hidden sector 
   is confined on the so-called ultraviolet (UV) brane.
In particular, quark superfields localize around the UV brane, while slepton superfields delocalize from the UV brane.
We will consider that the 5D MSSM can have a universal coupling between the MSSM multiplets and the hidden sector field.
However, because of this geometrical configuration of wave functions, we can naturally realize a large mass splitting 
   between squarks and sleptons.
Since gravitino, which is the superpartner of the massless 4D graviton, localizes around the UV brane, 
   its mass is large.  
Hence, the lightest neutralino serves as a dark matter candidate as usual in the MSSM. 

Taking into account a variety of experimental constraints, we find bulk field configurations 
   which are suitable to account for the muon $g-2$
   while reproducing the observed SM-like Higgs boson mass.

The structure of this paper is as follows. 
In Section 2, we describe the 5D MSSM with bulk superfields in the Randall-Sundrum background metric,  
    and derive the 4D MSSM as a low-energy effective theory. 
In Section 3, we consider various phenomenological constraints, and 
   present  the benchmark particle spectra of the model.
Section 4 is devoted for conclusion.

\section{
5D MSSM in the RS background metric
}



We consider a 5D space-time $(x^\mu, y)$, where the fifth dimension is compactified on the $S_1/Z_2$ orbifold,
  and $y$ defined in the range of $-\pi \leq y \leq \pi $ is an angle of $S_1$ with a radius $R$.
Because of $Z_2$ parity, $y$ is identified with $-y$, so that the orbifold possesses two fixed points at $y = 0$ and $y = \pi $.
Introducing two ``3-branes" located at these orbifold fixed points and assigning suitable brane tensions to them,
  a solution to the Einstein's equation is found to be \cite{Randall:1999ee}
\begin{eqnarray}
ds^2 &=& e^{-2R\sigma} \eta_{\mu\nu} dx^\mu dx^\nu - R^2 dy^2	,
\end{eqnarray}
where $\sigma = k|y|$, and $k$ is the AdS curvature.
The 4D Minkowski space is realized as a slice of  this AdS$_5$ space.

With this metric, a relation between the 5D Planck mass ($M_5$) and 
  the Planck mass ($M_4$) in the 4D effective theory is given by
\begin{eqnarray}
M_{4}^2	 =  	\frac{M_{5}^3}{k} \left( 1 - \omega^2 \right) 
\simeq \frac{M_5^3}{k},
\label{Planck}
\end{eqnarray} 
where $\omega = e^{-kR\pi}$ is the so-called warp factor, and we have assumed $\omega \ll 1$. 
In the following calculation, we simply take $k \simeq M_5 \simeq M_4$.
Because of the warped metric, an effective cutoff on the UV brane at $y=0$ is $M_{5}$ itself, 
   while that on the IR brane at $y=\pi$ is warped down to $M_{\rm cut} =M_{5} \, \omega$.
This effective cutoff at the IR brane is used to solve the gauge hierarchy problem 
   in the original paper by Randall and Sundrum \cite{Randall:1999ee}.


In this paper, we assume that all the MSSM superfields propagate in the 5D bulk, 
  while the SUSY breaking hidden sector is confined on the UV brane.
Since the gravity multiplet localizes toward the UV brane, the gravitino interacts strongly with the hidden sector field due to the large overlapping between their wave functions.
Therefore, when the SUSY is broken, the gravitino acquires a large mass.  
On the other hand, the MSSM gauge multiplets have flat configurations in the 5th dimensions, 
  and gaugino masses are suppressed by the volume of the 5th dimension. 
Hence the lightest neutralino is a dark matter candidate as usual in the MSSM. 
The wave function configuration for the MSSM matter multiplets depends on the bulk mass parameters. 
A sfermion mass becomes smaller, as its wave function is more delocalized from the UV brane.


Now we describe our $N=1$ 5D MSSM Lagrangian in terms of familiar ``4D-like" $N=1$ supermultiplets.\footnote{
In this paper, we assume some mechanism to stabilize the 5th dimensional radius, and 
  simply replace the radion supermultiplet as the 5th diminutional radius $R$. 
We refer Ref.~\cite{Maru:2003mq} for a very simple mechanism to stabilize the radion potential.  
}
In the Kaluza-Klein (KK) decomposition, only the $Z_2$-even 5D fields have massless modes.
After integrating over the extra 5th dimension $y$, these massless zero modes are identified 
  as usual 4D MSSM supermultiplets.



The 5D action for a bulk vector multiplet is described by using a 4D-like vector superfield
\begin{eqnarray}
V(x,y,\theta)
	&=&		-\theta \sigma^\mu \bar\theta A_\mu(x,y) 
			- i \bar\theta^2 \theta \lambda_1(x,y) 
			+ i \theta^2 \bar\theta \bar\lambda_1(x,y) 
			+ \frac{1}{2} \bar\theta^2 \theta^2 D(x,y) 	,
\end{eqnarray}
and a 4D-like adjoint chiral superfield
\begin{eqnarray}
\chi(x,y,\theta)
	&=&		\frac{1}{\sqrt{2}} (\Sigma(x,y) + i A_5(x,y)) 
			+ \sqrt{2} \theta \lambda_2(x,y) 
			+ \theta^2 F_\chi(x,y)	.
\end{eqnarray}
Under the $Z_2$ parity, the former is even while the latter is odd.
The gauge invariant 5D action for the gauge multiplet is given by
\begin{eqnarray}
S_5^{gauge}	&=&	\int d^4 x  \int_{-\pi}^\pi dy  
		\left\lbrace	
		\frac{1}{4g_5^2} \int d^2 \theta R \; \text{Tr} \left[W^\alpha W_\alpha \right] + \text{h.c.}
		\right. \nonumber	\\
& &		\left.
	+	\frac{2}{g_5^2} 
			\int d^4 \theta \frac{e^{-2R\sigma}}{2R} \text{Tr}
			\left[  
			\{ e^{V/2}, \partial_y e^{-V/2} \} +
			\frac{1}{\sqrt{2}} 
			(e^{V/2} \chi^\dagger e^{-V/2} + (e^{-V/2} \chi e^{V/2})
			\right]^2
		\right\rbrace	,
\end{eqnarray}
where the 5D gauge coupling $g_5$ has the mass dimension of $-1/2$.  
Since the wave function of the vector superfield zero-mode is found to be independent of the $y$-coordinate, 
  we have rescaled it as $V \to V/\sqrt{2 \pi R}$, by which the zero-mode is canonically normalized in the 4D effective theory 
  with the relation between the 5D and 4D gauge couplings, $g_5 = \sqrt{2 \pi R} \; g_4$.  


A hypermultiplet in the bulk is used to describe matter and Higgs multiplets,  
  which is decomposed into a pair of vector-like chiral superfields $\Phi$ and $\Phi^c$:
\begin{eqnarray}
\Phi(x,y,\theta)
	&=&	\phi(x,y) + \sqrt{2} \theta \psi(x,y) + \theta^2 F_\Phi(x,y) \; , \\
\Phi^c(x,y,\theta)
	&=&	\phi^c(x,y) + \sqrt{2} \theta \psi^c(x,y) + \theta^2 F_{\Phi^c}(x,y) \;.
\end{eqnarray}
Under the $Z_2$ parity, we assign an even parity for $\Phi$ while $\Phi^c$ is odd. 
The 5D action for the hypermultiplet is given by
\begin{eqnarray}
S_5^{matter}	&=& \int d^4 x  \int_{-\pi}^\pi dy 
	\left\lbrace
	\int d^4 \theta R e^{-2R\sigma} 
		\left( \Phi^\dagger e^{-V} \Phi + \Phi^c e^{V} \Phi^{c \dagger} \right)
	\right.	\nonumber \\
& & +
	\left.
	\int d^2 \theta e^{-3R\sigma} \Phi^c
	\left[ \partial_y - \frac{1}{\sqrt{2}} \chi 
		- \left( \frac{3}{2} - c_\Phi \right) R \sigma'
	\right] \Phi + \text{h.c.}
	\right\rbrace	,
\label{S5matter}
\end{eqnarray}
where $c_\Phi$ is a bulk mass parameter.
Due to the $Z_2$ parity assignment of $\Phi$ and $\Phi^c$, 
  only the $Z_2$-even chiral multiplet $\Phi$ has a zero-mode in the Kaluza-Klein decomposition.
By solving the SUSY vacuum condition, 
\begin{eqnarray}
\left[ \partial_y - \left( \frac{3}{2} - c_\Phi \right) R \sigma' \right] \Phi &=& 0,
\end{eqnarray}
we find the zero-mode wave function as 
\begin{eqnarray}
\Phi(x,y,\theta) \vert_\text{zero-mode}
	&=&	{\hat \Phi}(x,\theta) e^{(\frac{3}{2} - c_\Phi)R\sigma} 	.
\label{masslessZeroMode}
\end{eqnarray} 
Here, the 4D chiral superfiled ${\hat \Phi}(x,\theta)$ has a mass dimension $\frac{3}{2}$. 
The canonically normalized chiral superfield $\varphi_0(x,\theta)$ in the 4D effective theory is given by 
\begin{eqnarray}
{\hat \Phi}(x,\theta) = \sqrt{k} \mathcal{C}_\Phi \varphi_0(x,\theta)	\,	,
\label{redefine}
\end{eqnarray}
where $\varphi_0$ represents the 4D MSSM chiral superfields ($H_u$, $H_d$, $Q_i$, $U_i$, $D_i$, $L_i$, $E_i$), and	
\begin{eqnarray}
\mathcal{C}_\Phi = \sqrt{ \frac{(1-2c_\Phi)}{2 \left( \omega^{(-1+2c_\Phi)} -1 \right)}}	\quad	.
\end{eqnarray}
Note that the bulk mass parameter $c_\Phi$ controls the configuration of the zero-mode: 
  for $c_\Phi > 1/2$ ($c_\Phi < 1/2$), the zero-mode is localized toward the UV (IR) brane.\footnote{
In order to have the canonical Kahler potential in Eq.~(\ref{S5matter}), 
   we have redefined the hypermultiplet in Eq.~(\ref{masslessZeroMode}) as
   $\Phi \rightarrow \Phi e^{-R\sigma}$ when discussing about field localization in the extra dimension.
}

Now we introduce interaction terms among the bulk multiplets and a chiral multiplet 
   in the hidden sector on the UV brane.  
Because of the 5D $N=1$ SUSY, such interaction terms can be written only at the orbifold fixed points. 
In order to forbid phenomenologically dangerous terms such as $R$-parity violating terms, 
   we introduce an $R$-symmetry with the charge assignments listed in Table~\ref{R-charge}.
Here, a chiral superfield $X$ in the hidden sector has been introduced, 
  and  we assume that both the SUSY and the $R$-symmetry are broken by a vacuum expectation value (VEV) 
   of the $F$-component of $X$, $\langle F_X \rangle \neq 0$. %

\begin{table}
\begin{center}
\begin{math}
\begin{array}{|c||c|c|c|c|c|c|c|c||c|}
\hline
\text{Bulk field}  & V_a & Q_i^h & U_i^h & D_i^h &	 L_i^h &  E_i^h &  H_u^h & H_d^h & X  \\
\hline
R{\rm -charge}  & 0 & 1 &	1	&	1	&	1	&	1	&	0	&	0	&	0	\\
\hline 
\end{array}
\end{math}
\caption{
$R$-charge assignments for the 5D MSSM vector multiplet ($a = 1,2,3$), hypermultiplets and the hidden sector field $X$
(R-charge of $\theta$ is 1). 
Here, for example, $Q_i^h$ is a $Z_2$-even component of the bulk quark doublet hypermultiplet, 
  whose zero mode is identified as the quark doublet chiralsuperfield in the 4D MSSM. 
The generation index is denoted as $i=1,2,3$. 
}
\label{R-charge}
\end{center}
\end{table}


In the 5D MSSM, a Yukawa coupling is symbolically given by 
\begin{eqnarray}
S_5^{\rm Yukawa}	=
	\int d^4 x \int_{-\pi}^\pi dy \int d^2 \theta e^{-3R \sigma}
	\frac{1}{M_{5}^{3/2}} \Phi_1 \Phi_2 \Phi_3
	\left[
		Y_0 \, \delta(y) + Y_\pi \, \left\{ \delta(y+\pi) + \delta(y-\pi) \right\}
	\right]	\, ,
\label{5DYukawa}
\end{eqnarray}
where $Y_0$ and $Y_\pi$ are dimensionless coupling constants, 
   and $M_5^{3/2}$ is introduced to yield the correct mass dimension.
Here, $\Phi_1$ stands for the MSSM Higgs doublets, and the other two  $\Phi_2$ and $\Phi_3$ 
  stand for the MSSM matter multiplets. 
After the $y$-integration, a 4D effective Yukawa coupling is obtained as 
\begin{eqnarray}
Y_4  \simeq 
	\left[
		Y_0 + Y_\pi \, \omega^{-(\frac{3}{2} - c_{\Phi_1} - c_{\Phi_2} - c_{\Phi_3}) }
	\right]
	\mathcal{C}_{\Phi_1} \mathcal{C}_{\Phi_2} \mathcal{C}_{\Phi_3} \, , 
\end{eqnarray}
where we have used $k/M_5 \simeq 1$. 
Note that an appropriate choice of the bulk mass parameters can derive 
   an exponentially suppressed Yukawa coupling even for $Y_0,  Y_\pi ={\cal O}(1)$. 
Although this feature implies a possibility to naturally explain the Yukawa hierarchy in the SM, 
   in this paper we do not attempt to explain the Yukawa hierarchy, but concentrate on soft SUSY breaking parameters.


Let us consider $R$-symmetric contact terms between the hidden sector field $X$ and the 5D MSSM multiplets in the bulk. 
We introduce a contact term between the gauge multiplets and $X$ of the form:
\begin{eqnarray}
S_5^{Xg}
	=  \int d^4 x \int dy 
	\left\{ \int d^2 \theta \; d_a
	\frac{X}{M^2_5}  \text{Tr} \left[\tilde{W}^\alpha \tilde{W}_\alpha  \right]  + \text{h.c.}  \right\}
	\delta(y)	\, , 
\label{Xg5D}
\end{eqnarray}
where the original $\tilde{W}^\alpha$ has the mass dimension of $2$ before normalizing $V$. 
The contact terms between the Higgs hypermultiplets and the hidden sector field $X$ are written as 
\begin{eqnarray}
S_5^{Xh}
	&=& \int d^4 x \int dy  \int d^4 \theta
	\left\lbrace 
		\left[ 
		d_\mu \frac{X^\dagger}{M_{5}^2} H_u^h H_d^h + 
		d_{B_\mu} \frac{X^\dagger X}{M_{5}^3} H_u^h H_d^h + \text{h.c.}
		\right]
	\right.	\nonumber	\\
& &	\qquad\qquad	+
		\left[
		d_A^{H_u} \frac{X + X^\dagger}{M_{5}^2} H_u^{h\dagger} H_u^h +
		d_m^{H_u} \frac{X^\dagger X}{M_{5}^3} H_u^{h\dagger} H_u^h
		\right. \nonumber \\
& &	\qquad\qquad	+
	\left.
		\left.	
		d_A^{H_d} \frac{X + X^\dagger}{M_{5}^2} H_d^{h\dagger} H_d^h +
		d_m^{H_d} \frac{X^\dagger X}{M_{5}^3} H_d^{h\dagger} H_d^h
		\right]
	\right\rbrace
	\delta(y) \,	,
\label{Xh5D}
\end{eqnarray}
and those between the matter hypermultiplets and $X$ are
\begin{eqnarray}
S_5^{Xm}	
	= \int d^4 x \int dy 	\int d^4 \theta
	\left[
	(d_A^\Phi)_{ij} \frac{X + X^\dagger}{M_{5}^2} \Phi_i^\dagger \Phi_j +
	(d_m^\Phi)_{ij} \frac{X^\dagger X}{M_{5}^3} \Phi_i^\dagger \Phi_j
	\right] \delta(y)	\,	,
\label{Xm5D}
\end{eqnarray}
where $\Phi$ stands for $\{Q^h, U^h, D^h, L^h, E^h \}$ hypermultiplets, 
  and $\{i,j\}$ are generation indices. 
We can also introduce contact terms in the superpotential as follows: 
\begin{eqnarray}
S_5^{Xa}
	&=& \int d^4 x \int dy \int d^2 \theta
	\left\lbrace
	\frac{(a_u)_{ij}}{M_{5}^{5/2}} X H_u^h Q_i^h U_j^h +
	\frac{(a_d)_{ij}}{M_{5}^{5/2}} X H_d^h Q_i^h D_j^h 
	\right.	\nonumber \\
& &	\left. \qquad \qquad \qquad	+	\;
	\frac{(a_e)_{ij}}{M_{5}^{5/2}} X H_d^h L_i^h E_j^h + \text{h.c.} 
	\right\rbrace	\delta(y)	\,	.
\label{Xa5D}
\end{eqnarray}



The SUSY breaking by $\langle F_X \rangle$ induces the soft SUSY breaking terms 
  in the MSSM through the above contact terms at the effective 4D cutoff scale $M_{\rm cut} = M_4 \omega $. 
The gaugino masses are given by 
\begin{eqnarray}
M_a	& \simeq & - \sqrt{3} \left( \frac{d_a}{2 \pi R M_4 } \right) g_a^2 \; m_{3/2}	,	\quad	(a=1,2,3) \,	.
\label{gaugino_softmasses}
\end{eqnarray}
where $g_a$ is the SM gauge coupling, and the gravitino mass $m_{3/2}$ is given by
\begin{eqnarray}
m_{3/2} &=& \frac{\left< F_X \right>}{\sqrt{3} M_4} \quad . 
\end{eqnarray}
Note that the gaugino mass is suppressed by the so-called volume suppression factor of  $1/(2 \pi R M_4)$. 
Soft masses of the Higgs sector can be obtained from Eq.~(\ref{Xh5D}):
\begin{eqnarray}
m_{H_u}^2	&=&	
  3 \left[
  -	d_m^{H_u} 
  +	(d_A^{H_u})^2 \mathcal{C}_{H_u}^2 
  \right] \mathcal{C}_{H_u}^2 m_{3/2}^2
  \,	,	\\
m_{H_d}^2	&=&
  3 \left[
  - d_m^{H_d}  
  + (d_A^{H_d})^2 \mathcal{C}_{H_d}^2
  \right] \mathcal{C}_{H_d}^2 m_{3/2}^2
  \,	,	\\
B_\mu		&=&
	3 \, d_{B_\mu} \mathcal{C}_{H_u} \mathcal{C}_{H_d} m_{3/2}^2
-	
	\sqrt{3}
	\left(
	 d_A^{H_u} \mathcal{C}_{H_u}^2 
+	 d_A^{H_d} \mathcal{C}_{H_d}^2
	\right)  \mu \,	m_{3/2}   \, .
\label{Higgs_softmasses}
\end{eqnarray}
In our $R$-charge assignment, the $\mu$-term is forbidden, but it is generated through the SUSY breaking \cite{Giudice:1988yz}: 
\begin{eqnarray}
\mu		&=&
	\sqrt{3} \, d_\mu 
	\mathcal{C}_{H_u} \mathcal{C}_{H_d}	
	m_{3/2} \,	.
\label{Higgsmu}
\end{eqnarray}
Scalar soft masses of sparticles are generated from Eq.~(\ref{Xm5D}):
\begin{eqnarray}
(m_\Phi)_{ij}^2	&=&
	3 \left[
-	(d_m^\Phi)_{ij} 
	+	\sum_{n=1}^3
	(d_A^\Phi)_{in} (d_A^\Phi)_{nj} 
	\mathcal{C}_{\Phi_n}^2 
	\right] 
	\mathcal{C}_{\Phi_i} \mathcal{C}_{\Phi_j}   
	m_{3/2}^2	\,	,
\label{sparticle_softmasses}
\end{eqnarray}
where $\Phi$ stands for $Q, U, D, L, E$, and $i,j = \{1,2,3 \}$.
Last but not least, the trilinear couplings A-terms arise from Eq.~(\ref{Xa5D}):
\begin{eqnarray}
(A_u)_{ij}	&=&
	\frac{\sqrt{3} \, m_{3/2}}{(Y_u)_{ij}}	
	\left[
	(a_u)_{ij} \mathcal{C}_{H_u} \mathcal{C}_{Q_i} \mathcal{C}_{Q_j}
	\right.	\nonumber	\\
&&-	\left.	
	d_A^{H_u} (Y_u)_{ij} \mathcal{C}_{H_u}^2
-	\sum_{n=1}^3
	(d_A^Q)_{ni} (Y_u)_{nj} \mathcal{C}_{Q_n} \mathcal{C}_{Q_i}
-	\sum_{n=1}^3
	(d_A^U)_{nj} (Y_u)_{in} \mathcal{C}_{U_n} \mathcal{C}_{U_j}
	\right] 
	,		\\	
(A_d)_{ij}	&=&
	\frac{\sqrt{3} \, m_{3/2}}{(Y_d)_{ij}}	
	\left[
	(a_d)_{ij} \mathcal{C}_{H_d} \mathcal{C}_{Q_i} \mathcal{C}_{D_j}
	\right.	\nonumber	\\
&&-	\left.	
	d_A^{H_d} (Y_d)_{ij} \mathcal{C}_{H_d}^2
-	\sum_{n=1}^3
	(d_A^Q)_{ni} (Y_d)_{nj} \mathcal{C}_{Q_n} \mathcal{C}_{Q_i}
-	\sum_{n=1}^3
	(d_A^D)_{nj} (Y_d)_{in} \mathcal{C}_{D_n} \mathcal{C}_{D_j}
	\right] 
	,	\\	
(A_e)_{ij}	&=&
	\frac{\sqrt{3} \, m_{3/2}}{(Y_e)_{ij}}	
	\left[	
	(a_d)_{ij} \mathcal{C}_{H_d} \mathcal{C}_{L_i} \mathcal{C}_{E_j}
	\right.	\nonumber	\\
&&-	\left.	
	d_A^{H_d} (Y_e)_{ij} \mathcal{C}_{H_d}^2
-	\sum_{n=1}^3
	(d_A^L)_{ni} (Y_e)_{nj} \mathcal{C}_{L_n} \mathcal{C}_{L_i}
-	\sum_{n=1}^3
	(d_A^E)_{nj} (Y_e)_{in} \mathcal{C}_{E_n} \mathcal{C}_{E_j}
	\right] 
	\,	.	
\label{Aterms}
\end{eqnarray}
To avoid the SUSY flavor changing neutral currents (FCNCs), we assume 
  that the couplings $d_m^\Phi$, $d_A^\Phi$, $a_u$, $a_d$, $a_e$ are all flavor diagonal 
  and, in particular, flavor-universal for the first two generations.

\begin{figure}[ht]
\begin{center}
\scalebox{1.2}[1.2]{
\includegraphics[width=3.5in]{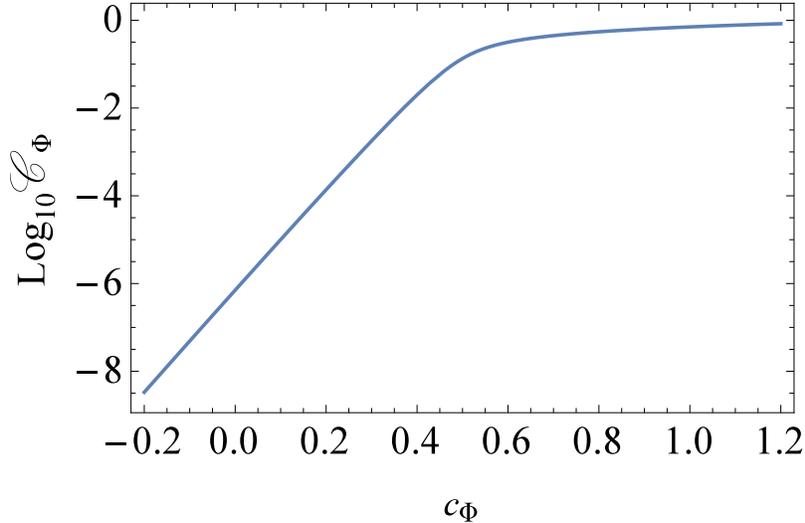}
}
\caption{Behavior of geometrical coefficient as a function of the bulk mass parameter:
$\mathcal{C}_\Phi = \sqrt{ \frac{(1-2c_\Phi)}{2 \left( \omega^{(-1+2c_\Phi)} -1 \right)}}$. The plot is a demonstration for the case with 
$\omega = 10^{-12}$.
}
\label{Geometrical_coefficient}
\end{center}
\end{figure}


All terms induced by the SUSY breaking are controlled by the gravitino mass  
  and the coupling constants for the contact interactions between the MSSM multiplets and  
  the hidden sector field $X$. 
In addition, the warped background geometry plays a crucial role in determining the size of the parameters. 
The gaugino masses are volume-suppressed from the gravitino mass. 
The scalar masses squared, the $A$-terms and the $\mu$-term  are controlled by the geometrical coefficients 
  of bulk hypermultiplets $\mathcal{C}_\Phi$.  
In Figure \ref{Geometrical_coefficient}, we show a geometrical coefficient $\mathcal{C}_\Phi$ as a function of 
  the bulk mass parameter $c_\Phi$. 
In this example, we have set $\omega=10^{-12}$ which generates the low cutoff scale of $\mathcal{O}(10^6)$ GeV.
\footnote{Basically, the choice of $\omega$ is arbitrary. But we have found that high cutoff scales result in tachyonic staus. Therefore, the choice of the cutoff scale of $\mathcal{O}(10^{6})$ GeV is preferable.}
For $c_\Phi=1/2$, the wave function is independent of $y$, 
   and $\mathcal{C}_\Phi ={\cal  O}(0.1)$ in this case corresponds to the volume suppression factor. 
As $c_\Phi >1/2$ increases, the wave function tends to localize towards the UV brane, 
    and hence $\mathcal{C}_\Phi$ is approaching to $1$. 
On the other hand, as $c_\Phi < 1/2$ decreases, the wave function tends to localize towards the IR brane,  
    and $\mathcal{C}_\Phi$ is being exponentially suppressed. 
Therefore, with a suitable choice of the bulk parameters, we can easily achieve a hierarchy 
    between soft SUSY breaking parameters. 
From Eqs.~(\ref{gaugino_softmasses})-(\ref{Aterms}), we see that the maximum values of the soft parameters are of 
  ${\cal O}(m_{3/2})$ when the coupling constants are of the order one. 
For $\omega \ll 1$, the gaugino masses are roughly an order of magnitude smaller than the gravitino mass. 
Thus, the model predicts the dark matter candidate to be the lightest neutralino.
We find that soft parameters for the scalars localized around the UV brane are of ${\cal O}(m_{3/2})$, 
  while they can be much smaller for the scalars localized around the IR brane. 
Interestingly, once all the couplings between the 5D MSSM multiplet and the hidden sector field are set to be universal, for instance of $\mathcal{O}(1)$, the diversity of soft masses and couplings of the 4D MSSM can be derived from the warped geometry with appropriate localization.

\section{
Benchmark particle mass spectra
}

In this section, we investigate realistic particle mass spectra which satisfy all phenomenological constraints. 
As we discussed in the previous section, a suitable choice of the bulk mass parameters can naturally generate a hierarchy 
   between sparticle masses.  
In the following analysis, we consider the 4D effective MSSM
 with the inputs of soft SUSY breaking parameters 
   at the 4D effective cutoff scale of $M_{\rm cut} \simeq  M_4 \omega$. 
The low energy mass spectrum is obtained through the renormalization group (RG) evolutions. 
We employ SOFTSUSY package (version 3.6.2) \cite{SOFTSUSY} to numerically solve the RG equations. 
With the output at low energies, other physical observables and constraints are computed 
  by using MicrOMEGAs package (version 4.2.3) \cite{MicrOMEGAs}.

Regarding to the inputs for the MSSM gaugino masses, we simply assume the universal couplings, $d_1=d_2=d_3$. 
Since the ratio $m_g = M_a/g_a^2 $ is RG invariant at the 1-loop level, the resultant mass ratio among 
  the gauginos is the same as those in the constrained MSSM when $m_g$ is set as a common input for the gaugino sector at the cutoff scale.
Hence the bino and wino are lighter than the gluino.
To avoid the severe experimental constraints on the SUSY FCNCs for the first two generations, 
   we assume that the couplings of the hidden sector with the first two generation matter fields are flavor-blind. 
In order to simplify our analysis, we set the universal soft mass inputs for the two Higgs doublets  ($m^0_h$),  
    the sleptons and squarks in the first two generations ($m^0_l$, $m^0_q$). 
The other free parameters in our analysis are the universal A-term $A_0$ at $M_{\rm cut}$ and $\tan\beta$.
We choose $\text{sign}(\mu) = +1$ to yield a positive $\Delta a_\mu$ 
  which can fill up the discrepancy between the experimental value and the SM prediction.

In our study, we consider various phenomenological constraints. 
We employ the combined result for the Higgs boson mass measured  by the ATLAS and the CMS collaborations \cite{mHiggs}. 
The lower mass bounds on squarks and gluino in the simplified model  \cite{gluino} are taken into account as a reference.  
As a motivation of this paper, the benchmark points are chosen such that the muon anomalous magnetic moment   
   $a_\mu = \frac{g_\mu - 2}{2}$ satisfies the current experimental value \cite{Exp_g-2, SM_g-2}.
Other constraints are from the branching ratios of rare decay processes: 
$b \rightarrow s + \gamma $			\cite{bsg},
$B_s \rightarrow \mu^+ + \mu^- $	\cite{Bsmumu},
$B \rightarrow \tau + \nu_\tau $	\cite{Btaunu},
$D_s \rightarrow \tau + \nu_\tau$	\cite{dtaunu-dmunu},
$D_s \rightarrow \mu + \nu_\mu$		\cite{dtaunu-dmunu},
and the Kaon decay parameter		\cite{Rl23}: 
\begin{equation}
R_{l23} = \left|	\frac{V_{us}(K_{l2})}{V_{us}(K_{l3})} \times
			\frac{V_{ud}(0^+ \rightarrow 0^+)}{V_{ud}(\pi_{l2})} \right|,
\end{equation}
where the CKM matrix elements, $V_{us}$ and $V_{ud}$, are measured 
  from the corresponding 3-body semileptonic Kaon decay ($K_{l3}$), 
  2-body leptonic Kaon and pion decay ($K_{l2}$, $\pi_{l2}$), 
  and super-allowed nuclear beta decay ($0^+ \rightarrow 0^+$).
The constraints which we employ are listed below:
\begin{eqnarray}
&&
m_h	= 125.09 \pm 0.21 (\text{stat.}) \pm 0.11 (\text{syst.}) \; \text{GeV},	
	\label{mH}	\\
&&
m_{\tilde{g}}	 \gtrsim 	1.4 \; \text{TeV} , 
 	\label{mgluino}	
	\\
&&
 \Delta a_\mu = a_\mu^{\text{exp}} - a_\mu^{\text{SM}} 
	= (28.6 \pm 8.0) \times 10^{-10} 		,		
	\label{gmu-2}\\
&&
 2.99 \times 10^{-4} < \text{BR}(b \rightarrow s + \gamma) < 3.87 \times 10^{-4},
 	\quad \quad (2 \sigma)	\\
&&
 2.1 \times 10^{-9} < \text{BR}(B_s \rightarrow \mu^+ + \mu^-) < 4.0 \times 10^{-9},
 	\quad (1 \sigma)	\\
&&
 0.15 < \frac{\text{BR}^{exp}(B_u \rightarrow \tau + \nu_\tau)}
	 {\text{BR}^{SM}(B_u \rightarrow \tau + \nu_\tau)} < 2.41	,
	 \qquad \qquad \qquad (3 \sigma) \\
&&
 5.07 \times 10^{-2} < \text{BR}(D_s \rightarrow \tau + \nu_\tau) < 6.03 \times
 		10^{-2}	,	\quad (2 \sigma)	\\
&& 
 5.31 \times 10^{-3} < \text{BR}(D_s \rightarrow \mu + \nu_\mu) < 5.81 \times 10^{-3} 	, 	\quad (1 \sigma)	\\
&&
 R_{l23} = 1.004 \pm 0.007  \; .
\end{eqnarray}
Since the LHC constraints require that sparticles must be heavy, their contributions to the precision electroweak observables are negligibly small \cite{RamseyMusolf:2006vr}.

Assuming R-parity conservation, the lightest neutralino is a primary candidate of the cold dark matter.
Beside the above constraints, we also consider the cosmological constraint 
  on the neutralino dark matter relic abundance.
Here we apply the result by the Planck satellite experiment \cite{PlanckCDM}:
\begin{eqnarray}
\Omega h^2	&=&	0.1188 \pm 0.0010 \quad (68\% \text{CL})  \; .
	\label{Omega}
\end{eqnarray}
Finally, the constraints from the results of the direct and indirect dark matter searches are taken into account. 
The most stringent upper limit on the spin-independent cross section of the neutralino dark matter 
   with nucleon has been reported by the LUX experiment \cite{LUX}, 
   while the IceCube experiment has set the most severe upper limit on the spin-dependent cross section 
   between the neutralino dark matter and nucleon \cite{IceCube}:
\begin{eqnarray}
\sigma_\text{SI}^{\chi-p} & \lesssim &
	7 \times 10^{-9} \; \text{pb} \; (90\% \; \text{CL}),	\quad 
	\text{for}	\quad 	m_\text{WIMP} \approx 600 \; \text{GeV},	\\
\sigma_\text{SD}^{\chi-p} & \lesssim &
	10^{-4}  \; \text{pb} \ (90\% \; \text{CL}),	\quad \quad \; \;
	\text{for}	\quad 	m_\text{WIMP} \approx 150-600 \; \text{GeV}.
\end{eqnarray}



The benchmark mass spectra along with the observables satisfying the above phenomenological constraints 
   are shown in Tables \ref{benchmark1}, \ref{benchmark2}, and \ref{benchmark3} 
   for three cutoff scales, $M_\text{cut} = 10^{5}, 10^{6}$ and $10^{7}$ GeV, respectively.
%
On Tables, $m_q^0 = 8500$ GeV is the common input soft mass for all squarks in the first two generations,
   while the common masses for the third generation sleptons and squarks are fixed to be $m_{l3}^0 = 800$ GeV 
   and $m_{q3}^0 = 9500$ GeV, respectively.
The choice of $m_{l3}^0$ for the benchmark point of the last column of Table~\ref{benchmark3} is a bit larger. 
The other four input parameters at the cutoff scale $\{ m_g, m^0_l, m^0_h, A_0\}$
   are chosen so as to satisfy the four most important constraints:  
   the Higgs boson mass,  gluino mass, 
   the muon anomalous magnetic dipole moment, and 
   the dark matter relic density.     
The mass hierarchy between squarks and sleptons/gauginos is crucial to reproduce $m_h \simeq 125$ GeV 
    and $\Delta a_\mu = \mathcal{O}(10^{-9})$ simultaneously.   
The  benchmark points satisfy all these phenomenological constraints. 



The dark matter neutralinos in the benchmarks are all bino-like.
%
Since we have chosen $m^0_{l3}$ larger than $m^0_l$ to avoid stau being the lightest sparticle (LSP),  
  the next-to-LSP (NLSP) is muon-sneutrino which almost degenerate with electron-sneutrino. 
The right dark matter relic abundance is achieved through the co-annihilation processes 
  between the neutralino LSP and the electron/muon-sneutrinos,
which is ensured by a correlation between the free inputs $m_g$ and $m^0_l$.
As can be seen in Eq.~(\ref{Delta-a_mu}), the sparticle contribution to the muon $g-2$ is proportional to $\tan\beta$. 
Hence, as the input of $\tan\beta$ is raised, the inputs of $m_g$ and $m^0_l$ are increased 
  to satisfy the constraint from the muon $g-2$.

The cutoff scales in all the tables are just about a few orders of magnitude higher than typical squark masses. 
Therefore, in our model, the distance of the RG evolutions of soft SUSY breaking parameters 
  are very short compared to, for example, the constrained MSSM, and hence 
  the RG evolution effects are much less. 
In addition, the inputs of the scalar squared masses are non-universal. 
In the slepton sector, we can see from the tables that the slepton masses at low energy 
   are smaller than the corresponding inputs at the boundary. 
This is the effect from the higher order corrections in the RG equations with the hierarchically large 
  inputs of the squark masses. 
Not as in the constrained MSSM, the left-handed sleptons of the first two generations 
  become lighter than right-handed ones. 
As the consequence, the NLSP in the provided spectra is muon-sneutrino.


\begin{table}
\begin{center}
\scalebox{1}[1]{
\begin{math}
\begin{array}{|c|cccc|}
\hline 

M_\text{cut} &       10^5 &       10^5 &       10^5 &       10^5 \\

       m_g &       1120 &       1178 &       1330 &       1615 \\

m^0_l, m^0_q &  306.0, 8500 &  312.1, 8500 &  331.0, 8500 &  370.1, 8500 \\

m^0_{l3}, m^0_{q3} &  800, 9500 &  800, 9500 &  800, 9500 &  800, 9500 \\

     m^0_h &       1000 &       2500 &       2600 &       2800 \\

       A_0 &      -6800 &      -2200 &      -1000 &          0 \\

 \tan\beta &         10 &         20 &         30 &         40 \\

\hline

       h^0 &     125.27 &     125.28 &     125.29 &     125.22 \\

  H^0, A^0 &       3716 &       3307 &       3057 &       2766 \\

     H^\pm &       3717 &       3308 &       3059 &       2767 \\

 \tilde{g} &       1405 &       1469 &       1639 &       1953 \\

\tilde{\chi}^0_{1,2} &  245, 484 &  257, 507 &  290, 571 &  352, 689 \\

\tilde{\chi}^0_{3,4} &  3600, 3601 &  2426, 2427 &  2278, 2279 &  2029, 2030 \\

\tilde{\chi}^\pm_{1,2} &  484, 3602 &  507, 2428 &  571, 2280 &  689, 2031 \\

\tilde{u},\tilde{c}_{L,R} &  8572, 8560 &  8573, 8561 &  8577, 8565 &  8585, 8573 \\

\tilde{d},\tilde{s}_{L,R} &  8572, 8559 &  8573, 8560 &  8577, 8564 &  8585, 8572 \\

\tilde{t}_{1,2} &  9018, 9317 &  9112, 9339 &  9124, 9317 &  9130, 9283 \\

\tilde{b}_{1,2} &  9307, 9555 &  9336, 9525 &  9316, 9474 &  9281, 9401 \\

\tilde{\nu}^{e,\mu}_L 	 &   253, 253 &  265, 265 &  298, 298 &  358, 358 \\

\tilde{e}, \tilde{\mu}_L &   265, 265 &  277, 277 &  308, 308 &  367, 367 \\

\tilde{e}, \tilde{\mu}_R &   335, 335 &  337, 337 &  354, 354 &  393, 392 \\

\tilde{\nu}^\tau_L &        765 &        766 &        760 &        751 \\

\tilde{\tau}_{1,2} &  726, 819 &  716, 827 &  677, 835 &  633, 833 \\

\hline 

\Delta a_\mu &  2.58 \times 10^{-9} &  2.94 \times 10^{-9} &  3.23 \times 10^{-9} &  2.62 \times 10^{-9} \\

\text{BR}(b \rightarrow s + \gamma) &    3.34 \times 10^{-4} &    3.34 \times 10^{-4} &    3.35 \times 10^{-4} &    3.36 \times 10^{-4} \\

\text{BR}(B_s \rightarrow \mu^+ + \mu^-) &    3.08 \times 10^{-9} &    3.07 \times 10^{-9} &    3.04 \times 10^{-9} &    2.98 \times 10^{-9} \\

\frac{\text{BR}^{exp}(B_u \rightarrow \tau + \nu_\tau)}
	 {\text{BR}^{SM}(B_u \rightarrow \tau + \nu_\tau)}
 	&  1.00 &  9.98 \times 10^{-1} &  9.95 \times 10^{-1} &  9.89 \times 10^{-1} \\

\text{BR}(D_s \rightarrow \tau + \nu_\tau) &    5.17 \times 10^{-2} &    5.17 \times 10^{-2} &    5.17 \times 10^{-2} &    5.17 \times 10^{-2} \\

\text{BR}(D_s \rightarrow \mu + \nu_\mu) &    5.33 \times 10^{-3} &    5.33 \times 10^{-3} &    5.33 \times 10^{-3} &    5.33 \times 10^{-3} \\

   R_{l23} &    1.000 &    1.000 &    1.000 &    1.000 \\

\hline

\Omega h^2 &  0.119 &  0.119 &  0.119 &  0.119 \\

\sigma_\text{SI}^{\chi-p} \; ({\rm pb}) &  9.35 \times 10^{-13} &  1.31 \times 10^{-12} &  1.46 \times 10^{-12} &  2.75 \times 10^{-12} \\

\sigma_\text{SD}^{\chi-p} \; ({\rm pb}) &  6.22 \times 10^{-10} &  3.88 \times 10^{-9} &  5.17 \times 10^{-9} &  8.74 \times 10^{-9} \\

\hline

\end{array}  
\end{math}}
\caption{
Benchmark particle mass spectra in GeV units for $M_\text{cut} = 10^5$ GeV.
Input soft masses for the first two generation squarks, 
the third generation slepton and squark are fixed as 
$m^0_q = 8500$ GeV,
$m^0_{l3} = 800$ GeV,
and $m^0_{q3} = 9500$ GeV. 
Other parameters including
the gaugino sector input $m_g$,
the soft masses for the first two generation sleptons $m^0_l$, and for two Higgs doublets $m^0_h$,
the universal trilinear coupling $A_0$,
and $\tan \beta$ are allowed to vary in this table.}
\label{benchmark1}
\end{center}
\end{table}


\begin{table}
\begin{center}
\scalebox{1}[1]{
\begin{math}
\begin{array}{|c|cccc|}
\hline

M_\text{cut} &       10^6 &       10^6 &       10^6 &       10^6 \\

       m_g &       1114 &       1300 &       1350 &       1400 \\

m^0_l, m^0_q &  349.4, 8500 &  365.3, 8500 &  370.5, 8500 &  374.0, 8500 \\

m^0_{l3}, m^0_{q3} &  800, 9500 &  800, 9500 &  800, 9500 &  800, 9500 \\

     m^0_h &       3200 &       3500 &       4000 &       4300 \\

       A_0 &      -6450 &      -2600 &      -2000 &       -400 \\

 \tan\beta &         10 &         20 &         30 &         40 \\

\hline

       h^0 &     125.00 &     125.16 &     125.21 &     125.22 \\

  H^0, A^0 &       4834 &       4343 &       4036 &       3541 \\

     H^\pm &       4835 &       4344 &       4036 &       3542 \\

 \tilde{g} &       1400 &       1606 &       1661 &       1715 \\

\tilde{\chi}^0_{1,2} &  243, 480 &  283, 557 &  294, 578 &  304, 596 \\

\tilde{\chi}^0_{3,4} &  3675, 3676 &  2945, 2946 &  2239, 2240 &  1558, 1560 \\

\tilde{\chi}^\pm_{1,2} &  480, 3676 &  558, 2947 &  578, 2241 &  596, 1561 \\

\tilde{u},\tilde{c}_{L,R} &  8557, 8546 &  8564, 8554 &  8566, 8556 &  8569(8), 8558 \\

\tilde{d}, \tilde{s}_{L,R} &  8557, 8546 &  8564, 8553 &  8567, 8555 &  8569, 8557 \\

\tilde{t}_{1,2} &  8666, 9130 &  8788, 9163 &  8784, 9113 &  8791, 9045 \\

\tilde{b}_{1,2} &  9125, 9535 &  9161, 9489 &  9111, 9396 &  9044, 9259 \\

\tilde{\nu}^{e,\mu}_L &  252, 252 &  291, 291 &  302, 301 &  312, 311 \\

\tilde{e}, \tilde{\mu}_L &   264, 264 &   302, 302 &   312, 312 &   322, 321 \\

\tilde{e},\tilde{\mu}_R &    368, 368 &   382, 381 &   386, 385 &   387, 386 \\

\tilde{\nu}^\tau_L &        736 &        729 &        680 &        631 \\

\tilde{\tau}_{1,2} &   699, 797 &   657, 799 &   554, 745 &   440, 681 \\

\hline

\Delta a_\mu &    2.31 \times 10^{-9} &    2.58 \times 10^{-9} &    2.88 \times 10^{-9} &    2.84 \times 10^{-9} \\

\text{BR}(b \rightarrow s + \gamma) &    3.33 \times 10^{-4} &    3.33 \times 10^{-4} &    3.33 \times 10^{-4} &    3.34 \times 10^{-4} \\

\text{BR}(B_s \rightarrow \mu^+ + \mu^-) &    3.08 \times 10^{-9} &    3.07 \times 10^{-9} &    3.06 \times 10^{-9} &    3.02 \times 10^{-9} \\

\frac{\text{BR}^{exp}(B_u \rightarrow \tau + \nu_\tau)}
	 {\text{BR}^{SM}(B_u \rightarrow \tau + \nu_\tau)}
	&  1.00 &  9.99 \times 10^{-1} &  9.97 \times 10^{-1} &  9.93 \times 10^{-1} \\

\text{BR}(D_s \rightarrow \tau + \nu_\tau) &    5.17 \times 10^{-2} &    5.17 \times 10^{-2} &    5.17 \times 10^{-2} &    5.17 \times 10^{-2} \\

\text{BR}(D_s \rightarrow \mu + \nu_\mu) &    5.33 \times 10^{-3} &    5.33 \times 10^{-3} &    5.33 \times 10^{-3} &    5.33 \times 10^{-3} \\

   R_{l23} &    1.000 &    1.000 &    1.000 &    1.000 \\

\hline

\Omega h^2 &    0.119 &    0.119 &    0.119 &    0.119 \\

\sigma_\text{SI}^{\chi-p} \; ({\rm pb})&   8.82 \times 10^{-13} &   7.78 \times 10^{-13} &   1.44 \times 10^{-12} &   4.96 \times 10^{-12} \\

\sigma_\text{SD}^{\chi-p} \; ({\rm pb}) &   5.65 \times 10^{-10} &    1.66 \times 10^{-9} &    5.59 \times 10^{-9} &    2.66 \times 10^{-8} \\

\hline

\end{array}  
\end{math}}
\caption{
Same as Table~\ref{benchmark1}, but for $M_\text{cut} = 10^6$ GeV.
}
\label{benchmark2}
\end{center}
\end{table}


\begin{table}
\begin{center}
\scalebox{1}[1]{
\begin{math}
\begin{array}{|c|cccc|}

\hline

M_\text{cut} &       10^7 &       10^7 &       10^7 &       10^7 \\

       m_g &       1280 &       1260 &       1230 &       1320 \\

m^0_l,m^0_q &  388.9, 8500 &  397.6, 8500 &  396.4, 8500 &  400.6, 8500 \\

m^0_{l3},m^0_{q3} &  800, 9500 &  800, 9500 &  800, 9500 &  1000, 9500 \\

     m^0_h &       1000 &       4500 &       5000 &       5100 \\

       A_0 &      -7500 &      -4000 &       -500 &          0 \\

 \tan\beta &         10 &         20 &         30 &         40 \\

\hline

       h^0 &     125.02 &     125.13 &     125.08 &     125.11 \\

  H^0, A^0 &       5611 &       5186 &       4723 &       4142 \\


     H^\pm &       5612 &       5187 &       4724 &       4143 \\

 \tilde{g} &       1589 &       1563 &       1524 &       1624 \\

\tilde{\chi}^0_{1,2} &  280, 551 &  275, 540 &  267, 524 &  286, 561 \\

\tilde{\chi}^0_{3,4} &  5548, 5548 &  3113, 3114 &  1966, 1968 &  1698, 1700 \\

\tilde{\chi}^\pm_{1,2} &  551, 5549 &  540, 3114 &  524, 1969 &  561, 1701 \\

\tilde{u}, \tilde{c}_{L,R} &  8553, 8544 &  8551, 8542 &  8549, 8541 &  8555(4), 8546 \\

\tilde{d}, \tilde{s}_{L,R} &  8553, 8542 &  8551, 8541 &  8550, 8540 &  8555, 8545(4) \\

\tilde{t}_{1,2} &  8377, 8984 &  8435, 8972 &  8476, 8926 &  8478, 8830 \\

\tilde{b}_{1,2} &  8979, 9525 &  8970, 9448 &  8924, 9323 &  8829, 9136 \\

\tilde{\nu}^{e,\mu}_L &  288, 288 &  283, 282 &  276, 275 &  295, 293 \\

\tilde{e}, \tilde{\mu}_L &  298, 298 &  294, 293 &  287, 286 &  306, 304 \\

\tilde{e}, \tilde{\mu}_R &  393, 392 &  404, 403 &  400, 399 &  403, 401 \\

\tilde{\nu}^\tau_L &        713 &        651 &        598 &        745 \\

\tilde{\tau}_{1,2} &  643, 788 &  536, 717 &  432, 648 &  471, 778 \\

\hline

\Delta a_\mu &    2.48 \times 10^{-9} &    2.69 \times 10^{-9} &    2.84 \times 10^{-9} &    3.13 \times 10^{-9} \\

\text{BR}(b \rightarrow s + \gamma) &    3.32 \times 10^{-4} &    3.32 \times 10^{-4} &    3.33 \times 10^{-4} &    3.33 \times 10^{-4} \\

\text{BR}(B_s \rightarrow \mu^+ + \mu^-) &    3.08 \times 10^{-9} &    3.08 \times 10^{-9} &    3.06 \times 10^{-9} &    3.03 \times 10^{-9} \\

\frac{\text{BR}^{exp}(B_u \rightarrow \tau + \nu_\tau)}
	 {\text{BR}^{SM}(B_u \rightarrow \tau + \nu_\tau)}
	&  1.00 &  9.99 \times 10^{-1} &  9.98 \times 10^{-1} &  9.95 \times 10^{-1} \\

\text{BR}(D_s \rightarrow \tau + \nu_\tau) &    5.17 \times 10^{-2} &    5.17 \times 10^{-2} &    5.17 \times 10^{-2} &    5.17 \times 10^{-2} \\

\text{BR}(D_s \rightarrow \mu + \nu_\mu) &    5.33 \times 10^{-3} &    5.33 \times 10^{-3} &    5.33 \times 10^{-3} &    5.33 \times 10^{-3} \\

   R_{l23} &    1.000 &    1.000 &    1.000 &    1.000 \\

\hline

\Omega h^2 &    0.119 &    0.119 &    0.119 &    0.119  \\

\sigma_\text{SI}^{\chi-p} \; ({\rm pb}) &   3.49 \times 10^{-13} &   6.34 \times 10^{-13} &   1.89 \times 10^{-12} &   3.17 \times 10^{-12} \\

\sigma_\text{SD}^{\chi-p} \; ({\rm pb}) &   5.81 \times 10^{-11} &    1.29 \times 10^{-9} &    9.70 \times 10^{-9} &    1.83 \times 10^{-8} \\

\hline

\end{array}  
\end{math}}
\caption{
Benchmark particle mass spectra in GeV units for $M_\text{cut} = 10^7$ GeV.
Input soft masses for the first two generation squarks, the third generation slepton and squark 
  are fixed as $m^0_q = 8500$ GeV, $m^0_{l3} = 800$ GeV, and $m^0_{q3} = 9500$ GeV.
In the last column,  the input for the 3rd generation slepton mass is taken a bit larger, $m^0_{l3} = 1000$ GeV. 
}
\label{benchmark3}
\end{center}
\end{table}


\section{Conclusions}

In order to reconcile the Higgs boson mass $m_h \simeq 125$ GeV 
   and the discrepancy of the muon anomalous magnetic dipole moment 
   $\Delta a_\mu \sim 10^{-9}$, 
   a hierarchical mass splittings between squarks and sleptons/gauginos are usually necessary.
In this paper, we have presented a 5D MSSM in the RS warped background metric 
   with the warp factor $\omega \ll 1$. 
All the MSSM multiplets reside in the bulk, while the SUSY is broken on the UV brane 
   where a hidden chiral field is localized. 
The zero-modes of the 5D MSSM fields are identified as the MSSM fields in the 4D effective theory.    
The SUSY breaking mediation to the MSSM sector is controlled by how much 
 the MSSM sparticles in the bulk overlap with the hidden field on the UV brane.  
Since the gravitino is localizing around the UV brane, the SUSY breaking parameters in the MSSM 
  are characterized by the gravitino mass $m_{3/2}$ and geometrical factors corresponding 
  to the zero-mode configurations.  
The gaugino masses are volume-suppressed $\sim 0.1 \; m_{3/2}$. 
Squarks are localized around the UV brane with a bulk mass parameter $> 1/2$,  
  while leptons acquire much smaller masses with a suitable choice of 
  the bulk mass parameter $< 1/2$.
Interestingly, assuming a common coupling between the hidden sector and the 5D MSSM sector, the diversity of the 4D MSSM soft terms can be derived from the universality of the underlying theory with the warped geometry and appropriate localization.
In our setup, a factor deference between bulk mass parameters, 
  which are the original parameters in the model, results in a hierarchy 
  because of the warped metric. 
With the hierarchical mass spectrum generated by the warped geometry, 
  we have demonstrated with the benchmarks that 
  not only $m_h \simeq 125$ GeV and $\Delta a_\mu \sim 10^{-9}$ can be reconciled, 
  but also various phenomenological constraints such as 
  the right abundance of the neturinalino dark matter,  
  the SUSY FCNC constraints and the LHC bounds on sparticle masses are all satisfied. 
In the benchmark points, squarks are too heavy to be produced at the LHC, 
  while sleptons, light charginos and neutralinos can be explored 
  at the LHC Run-2 in the future.

\section*{Acknowledgment}
H.M.T. would like to thank the Department of Physics and Astronomy 
  at the University of Alabama for hospitality during his visit. 
The work of N.O. is supported in part by the United States Department of Energy grant (DE-SC0013680).
The work of H.M.T. is supported in part by Vietnam National Foundation for Science 
  and Technology Development (NAFOSTED) under the grant No.~103.01-2014.22.


\end{document}